\documentclass[a4paper,12pt]{article}
\usepackage{latexsym}
\usepackage{amsmath}
\usepackage{amssymb}
\usepackage{amscd}

\newlength{\minitwocolumn}
\makeatletter
\@addtoreset{equation}{section}
\makeatother

\title{\bf 
The $SU(n)$ invariant massive Thirring model\\
with boundary reflection
}

\author{Takeo Kojima}
\date{\it
Department of Mathematics,
College of Science and Technology,\\
Nihon University, Chiyoda-ku, Tokyo
101-0062, Japan\\~\\
E-mail: kojima@math.cst.nihon-u.ac.jp\\~\\
{\rm \today}
}

\begin{document}
\maketitle

\begin{abstract}
We study the $SU(n)$ invariant massive Thirring model
with boundary reflection.
Our approach is based on the free field 
approach.
We construct
the free field realizations of the boundary state and it's
dual.
For an application of these realizations,
we present
integral representations for the form factors
of the local operators.
\end{abstract}

~\\

\section{Introduction}
For without-boundary massive theories,
integrability is ensured by the factorized scattering condition 
or the Yang-Baxter equation.
Cherdnik \cite{C} showed that the integrability in the presence
of reflecting boundary is ensured by the boundary Yang-Baxter
equation and the Yang-Baxter equation for without-boundary
theory.
A systematic treatment of integrable model with
boundary reflection was initiated by Sklyanin \cite{S}
in the framework of the Bethe Ansatz.
The boundary state condition was proposed
in \cite{GZ} on the basis of the boundary bootstrap
approach,
in order to obtain the boundary state.
Jimbo et al. \cite{JKKKM} developed
this idea to obtain the correlation functions
of the XXZ chain with a boundary,
using the free field approach \cite{JM}.
The $U_q(\widehat{sl_n})$ generalization of 
the XXZ chain with a boundary \cite{JKKKM},
was done in \cite{FK}.

In this paper we study the $SU(n)$ invariant massive Thirring
model with boundary reflection,
by means of the free field approach.
For $SU(2)$ symmetry case,
three men \cite{FKQ} studied the invariant massive
Thirring model with boundary reflection,
using the free field approach, and derived integral
representations for the form factors. 
In reference \cite{FKQ},their construction of
the boundary state starts with
Lukyanov's u-v cut off realization of
the Zamolodchikov-Faddeev operators \cite{L},
and the form factors are derived after removing the
cut-off parameter at the final stage.
In this paper we prefer to work directly
with operators with cut-off parameter removed.
We construct the free field realizations of the boundary
sate and it's dual, and give integral representations
for form factors of the $SU(n)$ invariant
massive Thirring model with a boundary.

Now a few words about the organization of this paper.
In section 2 we set up problem, and give the free field
realization of the Zamolodchikov-Faddeev operators.
In section 3 we construct the free field realizations
of the boundary state and it's dual.
In section 4 we give the free field
realization of the local operator.
and give integral representations of
the form factors of the local operators.
In Appendix we summarize useful facts about the multi-Gamma
functions.

\section{Formulation}

The purpose of this section is to set up the problem,
thereby fixing the notation.
We give the free field realization of the Zamolodchikov-
Faddeev operators.

\subsection{Model}

In this section we set up problem, and present briefly
necessary tools concering the completely
integral models of quantum field theory with
massive spectra.
The $SU(n)$ invariant massive Thirring model with
boundary reflection is described by the bulk $S$-matrix
and the boundary $S$-matrix ($K$-matrix).
Let $V$ be $n$-dimensional vector space $V=\oplus_{j=0}^{n-1}
{\mathbb C}v_j$.
The bulk $S$-matrix $S(\beta) \in {\rm End}(V \otimes V)$ 
of the present model
is given by
\begin{eqnarray}
S(\beta)=\frac{\Gamma\left(
\frac{n-1}{n}+\frac{\beta}{2\pi i}\right)\Gamma\left(
-\frac{\beta}{2\pi i}\right)}{
\Gamma\left(
\frac{n-1}{n}-\frac{\beta}{2\pi i}\right)\Gamma\left(
\frac{\beta}{2\pi i}\right)
}\times
\frac{\displaystyle
\beta-\frac{2\pi i}{n}P}{
\displaystyle
\beta-\frac{2\pi i}{n}},
\end{eqnarray}
where the operator $P \in {\rm End}(V \otimes V)$ 
represents the permutation.
\begin{eqnarray}
P(a \otimes b)=b \otimes a.
\end{eqnarray}
This $S$-matrix $S(\beta)$ satisfies the Yang-Baxter
equation.
\begin{eqnarray}
S_{12}(\beta_1-\beta_2)S_{13}(\beta_1-\beta_3)
S_{23}(\beta_2-\beta_3)=
S_{23}(\beta_2-\beta_3)S_{13}(\beta_1-\beta_3)
S_{12}(\beta_1-\beta_2).
\end{eqnarray}
Let us fix two numbers $L,M$
such that
$
0\leq L <M \leq n,~~(L,M \in {\mathbb{N}}).
$
The boundary $K$-matrix $K(\beta)\in {\rm End}(V)$ of the
present model is a diagonal matrix given by
\begin{eqnarray}
K(\beta)v_j=\sum_{j=0}^{n-1}v_k \delta_{j,k} K(\beta)_j^j.
\end{eqnarray}
Here the diagonal elements are defined by
\begin{eqnarray}
K(\beta)_j^j=k_{L,M}(\beta)\left\{
\begin{array}{cc}
1,& 0\leq j \leq L-1,\\
\frac{\displaystyle \mu-\beta}{
\displaystyle \mu+\beta},& L\leq j \leq M-1,\\
1,& M\leq j \leq n-1.
\end{array}
\right.
\end{eqnarray}
Here the scalar function $k_{L,M}(\beta)$ is given by
\begin{eqnarray}
k_{L,M}(\beta)&=&
\frac{\Gamma\left(\frac{-\beta+\mu}{2\pi i}\right)
\Gamma\left(\frac{\beta+\mu}{2\pi i}+1-\frac{L}{n}\right)
}{
\Gamma\left(\frac{\beta+\mu}{2\pi i}\right)
\Gamma\left(\frac{-\beta+\mu}{2\pi i}+1-\frac{L}{n}\right)
}
\times
\frac{\Gamma\left(\frac{-\beta-\mu}{2\pi i}
+\frac{L}{n}\right)
\Gamma\left(\frac{\beta-\mu}{2\pi i}+1+\frac{L-M}{n}\right)
}{
\Gamma\left(\frac{\beta-\mu}{2\pi i}
+\frac{L}{n}\right)
\Gamma\left(\frac{-\beta-\mu}{2\pi i}+1+\frac{L-M}{n}\right)
}
\nonumber\\
&\times&
\frac{
\Gamma\left(
\frac{-\beta}{2\pi i}+\frac{1}{2}-\frac{1}{2n}\right)
\Gamma\left(
\frac{\beta}{2\pi i}
\right)
}{
\Gamma\left(
\frac{\beta}{2\pi i}+\frac{1}{2}-\frac{1}{2n}\right)
\Gamma\left(
\frac{-\beta}{2\pi i}
\right)
}.\nonumber\\
\end{eqnarray}
The $S$-matrix and the boundary $K$-matrix satisfy
the boundary Yang-Baxter equation.
\begin{eqnarray}
K_2(\beta_2)S_{2 1}(\beta_1+\beta_2)
K_1(\beta_1)S_{1 2}(\beta_1-\beta_2)\nonumber\\
=S_{21}(\beta_1-\beta_2)
K_1(\beta_1)S_{1 2}(\beta_1+\beta_2)
K_2(\beta_2).
\end{eqnarray}
For the description of the space of physical states
we use the Zamolodchikov-Faddeev operators.
The Zamolodchikov-Faddeev operators 
$Z_j^*(\beta),
Z_j(\beta),~(j=1,\cdots,n-1)$
of the present model
satisfy the following commutation relations.
\begin{eqnarray}
Z_{j_1}^*(\beta_1)Z_{j_2}^*(\beta_2)&=&\sum_{k_1,k_2=0}^{n-1}
S(\beta_1-\beta_2)_{j_1 j_2}^{k_1 k_2}
Z_{k_2}^*(\beta_2)Z_{k_1}^*(\beta_1),\label{Com1}
\\
Z_{j_1}(\beta_1)Z_{j_2}(\beta_2)&=&
\sum_{k_1,k_2=0}^{n-1}
S(\beta_1-\beta_2)_{k_1 k_2}^{j_1 j_2}
Z_{k_2}(\beta_2)Z_{k_1}(\beta_1).\label{Com2}
\end{eqnarray}
Here $S(\beta)_{ab}^{cd}$
are matrix elements of the $S$-matrix,
\begin{eqnarray}
S(\beta)v_{k_1}\otimes v_{k_2}=
\sum_{j_1,j_2=0}^{n-1}v_{j_1}\otimes v_{j_2}
S(\beta)_{j_1 j_2}^{k_1 k_2}.
\end{eqnarray}
The Zamolodchikov-Faddeev operators
$Z_j^*(\beta)$ and it's dual $Z_j(\beta)$ satisfy
the inversion relation.
\begin{eqnarray}
Z_j^*(\beta_1)Z_k(\beta_2+\pi i)=\frac{\delta_{j,k}}
{\beta_1-\beta_2}+
\cdots,~~(\beta_1 \to \beta_2).\label{Inversion}
\end{eqnarray}
where ``$\cdots$'' means regular term.
\\
For the description of the space of physical state
we use the boundary state $|B\rangle$. \cite{GZ}\\
The boundary state 
$|B\rangle$ and it's dual $\langle B|$ are characterized by
the following conditions.
\begin{eqnarray}
K(\beta)_j^j
Z_j^*(\beta)|B\rangle=
Z_j^*(-\beta)|B\rangle,~(j=0,\cdots,n-1),
\\
K(\beta)_j^j\langle B|Z_j(-\beta+\pi i)=
\langle B|Z_j(\beta+\pi i),~(j=0,\cdots,n-1).
\end{eqnarray}
In this paper we shall construct the free field 
realizations of the boundary state and it's dual.

The space of states is generated by
the vectors,
\begin{eqnarray}
|\beta_1,\cdots,\beta_N \rangle_{j_1,\cdots, j_N}=
Z_{j_1}^*(\beta_1)\cdots Z_{j_N}^*(\beta_N)|B\rangle,
\end{eqnarray}
and the dual space,
\begin{eqnarray}
~^{j_1,\cdots, j_N}
\langle \beta_1,\cdots,\beta_N|=
\langle B|Z_{j_1}(\beta_1)\cdots Z_{j_N}(\beta_N).
\end{eqnarray}
Consider local operators $
{\cal O}$ and construct the matrix element,
\begin{eqnarray}
~^{j_1\cdots j_M}\langle \beta_1' \cdots \beta_M'|
{\cal O}|\beta_1 \cdots \beta_N \rangle_{k_1 \cdots k_N}.
\end{eqnarray}
We call the above matrix elements ``form factor''.

In this paper we shall give integral representations
of the form factors.

In this subsection we have introduced
the basic tools of the bootstrap approach, i.e.
the $S$-matrix $S(\beta)$, the boundary $K$-matrix $K(\beta)$,
the Zamolodchikov-Faddeev operators 
$Z_j^*(\beta),~Z_j(\beta)$,
the boundary state $|B\rangle$ and it's dual state
$\langle B|$,
the basis of the space of the states ;
$|\beta_1,\cdots,\beta_N \rangle_{j_1,\cdots, j_N}
$,
it's dual :
$~^{j_1,\cdots, j_N}
\langle \beta_1,\cdots,\beta_N|
$, and the form factors.

\subsection{Zamolodchikov-Faddeev operators}

The purpose of this section is to
give the free field realization of
the Zamolodchikov-Faddeev operators.\\
Let $a_j(t)~(1\leq j \leq n-1, t \in {\mathbb{R}})$
be the free bose field satisfying the
following commutation relation,
\begin{eqnarray}
[a_j(t),a_k(t')]=-\frac{1}{t}\frac{{\rm sh}\frac{(a_j|a_k)\pi
t}{n}}{
{\rm sh}\frac{\pi
t}{n}}e^{\frac{\pi}{n}|t|}
\delta(t+t').
\end{eqnarray}
Here $((a_j|a_k))_{1\leq j,k \leq n-1}$
is the Cartan matrix of type $A_{n-1}$.

\begin{eqnarray}
((a_j|a_k))_{1\leq j,k \leq n-1}=
\left(
\begin{array}{cccccc}
2&-1&0&\cdots&\cdots&0\\
-1&2&-1&0\cdots&\cdots\\
0&-1&2&-1&0&\cdots\\
\cdots&\cdots&\cdots&\cdots&\cdots&\cdots
\\
\cdots&\cdots&0&-1&2&-1
\\
0&\cdots&\cdots&0&-1&2
\end{array}\right).
\end{eqnarray}

~\\
Let us introduce the Fock space ${\cal H}$ generated by the vacuum
vector $|vac\rangle$ satisfying
\begin{eqnarray}
a_j(t)|vac\rangle =0~{\rm for}~t>0.
\end{eqnarray}
A normal ordering $:A:$ of an element $A$ is defined as
usual :
anihilation operators are replaced on the right of the
creation operators, for example,
\begin{eqnarray}
:a(t_1)a(-t_2):=a(-t_2)a(t_1),~(t_1,t_2>0).
\end{eqnarray}
Let us introduce the basic operators,
\begin{eqnarray}
V_j(\beta)&=&:\exp\left(
\int_{-\infty}^\infty
a_j(t)e^{i\beta t}dt\right):~(1\leq j \leq n-1),\\
V_0(\beta)&=&:\exp\left(
\int_{-\infty}^\infty
a_1^*(t)e^{i\beta t}dt\right):,\\
V_{n}(\beta)&=&:\exp\left(
\int_{-\infty}^\infty
a_{n-1}^*(t)e^{i\beta t}dt\right):,
\end{eqnarray}
where we have set
\begin{eqnarray}
a_1^*(t)&=&-\sum_{j=1}^{n-1}a_j(t)\frac{{\rm sh}
\frac{(n-j)\pi t}{n}}{{\rm sh}\pi t},\\
a_{n-1}^*(t)&=&-\sum_{j=1}^{n-1}a_j(t)\frac{{\rm sh}
\frac{j \pi t}{n}}{{\rm sh}\pi t}.
\end{eqnarray}
We have
\begin{eqnarray}
[a_1^*(t),a_j(t')]&=&\delta_{1,j}\frac{e^{\frac{\pi}{n}|t|}}{t}
\delta(t+t'),
\end{eqnarray}
\begin{eqnarray}
[a_j(t),a_{n-1}^*(t')]&=&\delta_{j,n-1}
\frac{e^{\frac{\pi}{n}|t|}}{t}
\delta(t+t').
\end{eqnarray}
The basic operators satisfy the following
contraction relations.
\begin{eqnarray}
V_j(\beta_1)V_{j-1}(\beta_2)=
\frac{-ie^{-\gamma}}{\beta_2-\beta_1+\frac{\pi i}{n}}
:V_j(\beta_1)V_{j-1}(\beta_2)
:,
~(j=1,\cdots, n),\\
V_{j-1}(\beta_1)V_j(\beta_2)=
\frac{-ie^{-\gamma}}{\beta_2-\beta_1+\frac{\pi i}{n}}
:V_{j-1}(\beta_1)V_j(\beta_2):,~
(j=1,\cdots, n),
\\
V_j(\beta_1)V_j(\beta_2)=
e^{2\gamma}(\beta_1-\beta_2)
\left(
\beta_2-\beta_1+\frac{2\pi i}{n}
\right)
:V_j(\beta_1)V_j(\beta_2):,
~(j=1,\cdots,n-1),
\end{eqnarray}
\begin{eqnarray}
V_0(\beta_1)V_0(\beta_2)=e^{\frac{n-1}{n}
(\gamma+{\rm log}2\pi)}
\frac{\Gamma\left(\frac{\beta_1-\beta_2}{2\pi i}+1-\frac{1}{n}
\right)}{
\Gamma\left(\frac{\beta_1-\beta_2}{2\pi i}\right)}:
V_0(\beta_1)V_0(\beta_2):\\
V_n(\beta_1)V_n(\beta_2)=
e^{\frac{n-1}{n}
(\gamma+{\rm log}2\pi)}
\frac{\Gamma\left(\frac{\beta_1-\beta_2}{2\pi i}+1-\frac{1}{n}
\right)}{
\Gamma\left(\frac{\beta_1-\beta_2}{2\pi i}\right)}
:V_n(\beta_1)V_n(\beta_2):.
\end{eqnarray}
The basic operators satisfy the following commutation
relations.
\begin{eqnarray}
V_j(\beta_1)V_{j-1}(\beta_2)&=&
\frac{\beta_1-\beta_2+\frac{\pi i}{n}}{
\beta_2-\beta_1+\frac{\pi i}{n}
}V_{j-1}(\beta_2)V_j(\beta_1),~(j=1,\cdots,n),\\
V_j(\beta_1)V_{j}(\beta_2)&=&-
\frac{\beta_1-\beta_2-\frac{2 \pi i}{n}}{
\beta_2-\beta_1-\frac{2 \pi i}{n}
}V_{j}(\beta_2)V_j(\beta_1),~(j=1,\cdots,n-1),
\end{eqnarray}
and
\begin{eqnarray}
V_0(\beta_1)V_0(\beta_2)&=&s(\beta_1-\beta_2)
V_0(\beta_2)V_0(\beta_1),\\
V_{n}(\beta_1)V_n(\beta_2)&=&s(\beta_1-\beta_2)
V_n(\beta_2)V_n(\beta_1).
\end{eqnarray}
Here we have set 
\begin{eqnarray}
s(\beta)=
\frac{\Gamma\left(\frac{\beta}{2\pi i}+1-\frac{1}{n}\right)
\Gamma\left(
-\frac{\beta}{2\pi i}
\right)}{
\Gamma\left(-\frac{\beta}{2\pi i}+1-\frac{1}{n}\right)
\Gamma\left(
\frac{\beta}{2\pi i}
\right)
}.
\end{eqnarray}

~\\

Next we give the free field realization of the Zamolodchikov-
Faddeev operators.\\
Let us set the operators $Z_j^*(\beta), (j=0,\cdots,n-1)$ by
\begin{eqnarray}
Z_j^*(\beta)&=&\int_{-\infty}^\infty d\alpha_1 \cdots 
\int_{-\infty}^\infty
d\alpha_j
\frac{V_0(\beta)V_1(\alpha_1) \cdots V_j(\alpha_j)}
{\prod_{l=1}^j (\alpha_{l-1}-\alpha_l+\frac{\pi i}{n})}
\nonumber
\\
&=&
(-ie^{-\gamma})^j
\int_{-\infty}^\infty d\alpha_1 \cdots 
\int_{-\infty}^\infty
d\alpha_j\nonumber\\
&\times&
:V_0(\beta)V_1(\alpha_1) \cdots V_j(\alpha_j):
\prod_{l=1}^j\frac{1}{
\left(\alpha_l-\alpha_{l-1}+\frac{\pi i}{n}\right)
\left(\alpha_{l-1}-\alpha_{l}+\frac{\pi i}{n}\right)
}
, 
\end{eqnarray} 
where we set $\alpha_0=\beta$.\\
Let us set the operators $Z_j(\beta),~(j=0,\cdots, n-1)$ by
\begin{eqnarray}
Z_j(\beta)&=&c_j\int_{-\infty}^\infty d\alpha_{j+1}
\cdots \int_{-\infty}^\infty d\alpha_{n-1}
\frac{V_{j+1}(\alpha_{j+1})\cdots
V_{n-1}(\alpha_{n-1})V_{n}(\beta)}{
\prod_{l=j+1}^{n-1}(\alpha_l-\alpha_{l+1}+\frac{\pi i}{n})}
\nonumber\\
&=&(-ie^{-\gamma})^{n-j-1}c_j
\int_{-\infty}^\infty d\alpha_{j+1} \cdots 
\int_{-\infty}^\infty
d\alpha_{n-1}\nonumber\\
&\times&
:V_{j+1}(\alpha_{j+1})
\cdots V_{n-1}(\alpha_{n-1})V_n(\beta):
\prod_{l=j+1}^{n-1}\frac{1}{
\left(\alpha_l-\alpha_{l+1}+\frac{\pi i}{n}\right)
\left(\alpha_{l+1}-\alpha_{l}+\frac{\pi i}{n}\right)
}
, \nonumber\\
\end{eqnarray}
where we set $\alpha_n=\beta$, and 
$c_j$ are proper constants choosen to satisfy the relation
(\ref{Inversion}).\\
The operators $Z_j^*(\beta), Z_j(\beta), (j=0,\cdots, n-1,
\beta \in {\mathbb{R}})$ generate the Zamolodchikov-Faddeev 
algebras.
The above free field realizations of 
the Zamolodchikov-Faddeev operators $Z_j^*(\beta),
Z_j(\beta)$ satisfy the commutation relations
(\ref{Com1}), (\ref{Com2}), and the inversion
relation (\ref{Inversion}).
The proof is as the same as those in the reference \cite{KY}.

~\\
{\it Note.~
The free field realization of
the Zamolodchikov-Faddeev operators $Z_j^*(\beta)$ 
of the $A_{n-1}^{(1)}$ Toda fields thory are given in
the reference \cite{MT}.
Naively the Zamolodchikov-Faddeev operators
(the type-II vertex operators)
of the present model is a limiting case of those of
the $A_{n-1}^{(1)}$
Toda field theory.
The construction of the dual 
Zamolodchikov-Faddeev operators $Z_j(\beta)$
is skeched in the reference \cite{KY}.
In the reference \cite{KY}
they constructed
the type-I vertex operators and it's dual vertex operators
for the $A_{n-1}^{(1)}$ critical chain.
}

\section{Boundary State}

In this section we give the free field realization
of the boundary state $|B\rangle$,
which satisfies 
\begin{eqnarray}
K(\beta)_j^j Z_j^*(\beta)|B\rangle=
Z_j^*(-\beta)|B\rangle,~~(j=0,\cdots,n-1).
\label{character}
\end{eqnarray}
Here $K(\beta)_j^j$ are diagonal elements of
the boundary $K$-matrix.\\
We make the ansatz that the boundary state has the following
form.
\begin{eqnarray}
|B\rangle=e^F|vac\rangle.
\end{eqnarray}
Here $F$ is a quadratic form of the free bosons,
\begin{eqnarray}
F=\sum_{j,k=1}^{n-1}\int_0^\infty
\alpha_{j,k}(t)a_j(-t)a_k(-t) dt +
\sum_{j=1}^{n-1}\int_0^\infty
\beta_j(t)a_j(-t)dt,\label{def:F}
\end{eqnarray}
where $\alpha_{j,k}(t)$ and $\beta_j(t)$ are scalar functions.\\
In this section we show that
we can choose the coefficients $\alpha_{j,k}(t)$ and
$\beta_j(t)$ suth that the state $|B\rangle$ satisfies
the characterizing relation (\ref{character}).

The operator $e^F$ has the effect of a Bogoliubov transformation,
\begin{eqnarray}
e^{-F}a_l(-t)e^F&=&0,~~(t>0),
\end{eqnarray}
\begin{eqnarray}
e^{-F}a_l(t)e^{F}&=&a_l(t)+\sum_{j,k=1}^{n-1}
\alpha_{j,k}(t)\left(-\frac{e^{\frac{\pi}{n}t}}{t}
\frac{{\rm sh}\frac{(a_l|a_j)\pi t}{n}}{
{\rm sh}\frac{\pi t}{n}
}a_k(-t)
-\frac{e^{\frac{\pi}{n}t}}{t}
\frac{{\rm sh}\frac{(a_l|a_k)\pi t}{n}}{
{\rm sh}\frac{\pi t}{n}
}a_j(-t)
\right)\nonumber\\
&-&\sum_{l=1}^{n-1}\beta_l(t)
\frac{e^{\frac{\pi}{n}t}}{t}
\frac{{\rm sh}\frac{(a_l|a_j)\pi t}{n}}{
{\rm sh}\frac{\pi t}{n}
}
,~~(t>0).
\end{eqnarray}
Let us set the coefficient $\alpha_{j,k}(t)$ by
\begin{eqnarray}
\alpha_{j,k}(t)=-\frac{t e^{-\frac{\pi}{n}t}}{2}\times
I_{j,k}(t).\label{def:alpha}
\end{eqnarray}
Here $(I_{j,k}(t))_{1\leq j,k \leq n-1}$
is the inverse matrix of the quantum Cartan matrix of type 
$A_{n-1}$.
The quantum Cartan matrix of type $A_{n-1}$ is defined by
\begin{eqnarray}
\left(
\frac{
\displaystyle
{\rm sh}\frac{(a_j|a_k)\pi t}{n}}
{
\displaystyle
{\rm sh}\frac{\pi t}{n}
}\right)_{1\leq j,k \leq n-1},
\end{eqnarray}
where $((a_j|a_k))_{1\leq j,k \leq n-1}$
is the Cartan matrix of type $A_{n-1}$.\\
Explicitly matrix elements $I_{j,k}(t)$ of the inverse matrix
of the quantum Cartan matrix are given by
\begin{eqnarray}
I_{j,k}(t)=
\frac{\displaystyle
{\rm sh}\frac{j \pi t}{n}~
{\rm sh}\frac{(n-k) \pi t}{n}
}{
\displaystyle
{\rm sh} \frac{\pi t}{n}~
{\rm sh} \pi t 
}=I_{k,j}(t),~(1\leq j \leq k \leq n-1).
\label{def:Inverse}
\end{eqnarray}
After specifying $\alpha_{j,k}(t)$, we have
\begin{eqnarray}
e^{-F}a_l(t)e^{F}=a_l(t)+a_l(-t)
-\frac{e^{\frac{\pi}{n}t}}{
t}~\sum_{j=1}^{n-1}
\beta_j(t)\times\frac{
{\rm sh}\frac{(a_j|a_l)\pi t}{n}}
{
{\rm sh}\frac{\pi t}{n}
},~~(t>0).
\end{eqnarray}
In what follows we use the abbreviations,
\begin{eqnarray}
V_j^a(\beta)=
\exp\left(\int_0^\infty
a_j(t)e^{i\beta t} dt\right),\\
V_j^c(\beta)=
\exp\left(\int_0^\infty
a_j(-t)e^{-i\beta t} dt\right),
\end{eqnarray}
\begin{eqnarray}
V_0^a(\beta)=
\exp\left(\int_0^\infty
a_1^*(t)e^{i\beta t} dt\right),
\\
V_0^c(\beta)=
\exp\left(\int_0^\infty
a_{1}^*(-t)e^{-i\beta t} dt\right),
\end{eqnarray}
\begin{eqnarray}
V_n^a(\beta)=
\exp\left(\int_0^\infty
a_{n-1}^*(t)e^{i\beta t} dt\right),
\\
V_n^c(\beta)=
\exp\left(\int_0^\infty
a_{n-1}^*(-t)e^{-i\beta t} dt\right).
\end{eqnarray}
The actions of the basic operators on the vector
$|B\rangle$ are given by
\begin{eqnarray}
V_j^a(\alpha)|B\rangle&=&
G_j(\alpha)V_j^c(-\alpha)|B\rangle,~(j=1,\cdots,n-1),\\
V_0^a(\beta)|B\rangle&=&
H_0(\beta)V_0^c(-\beta)|B\rangle.
\end{eqnarray}
Here the functions $G_j(\alpha),~(j=1,\cdots,n-1)$ and
$H_0(\beta)$ are given by
\begin{eqnarray}
G_j(\alpha)&=&\exp\left(-
\frac{1}{2}
\int_0^\infty
\frac{e^{2i\alpha t}}{t}(1+e^{\frac{2\pi t}{n}})dt
-\int_0^\infty
\frac{e^{(i \alpha +\frac{\pi}{n})t}}{t}
\sum_{l=1}^{n-1}\beta_l(t)\frac{{\rm sh}\frac{(a_l|a_j)\pi t}{n}}
{{\rm sh}\frac{\pi t}{n}}dt
\right),\nonumber
\\
\\
H_0(\beta)&=&\exp\left(
-\frac{1}{2}\int_0^\infty
\frac{e^{(2i\beta+\frac{\pi}{n})t}}{t}
\frac{{\rm sh}\frac{(n-1)}{n}\pi t}{
{\rm sh}\pi t}dt
+
\int_0^\infty
\frac{e^{(i\beta+\frac{\pi}{n})t}}{t}
\beta_1(t)dt
\right).\nonumber\\
\end{eqnarray}
Let us set the coefficients $\beta_l(t),~(l=1,\cdots,n-1)$ by
\begin{eqnarray}
\beta_l(t)=
\frac{1}{2}(e^{-\frac{\pi}{n}t}-1)\sum_{k=1}^{n-1}I_{k,l}(t)
+e^{i\mu t+\frac{\pi}{n}(L-1)t}I_{L,l}(t)
+e^{-i\mu t+\frac{\pi}{n}(M-2L-1)t}I_{M,l}(t).
\label{def:beta}
\end{eqnarray}
Here we should understand
$I_{0,l}(t)=0$ and
$I_{n,l}(t)=0$.\\
After specifying the coefficients 
$\beta_l(t)$, we have
\begin{eqnarray}
e^{-F}a_l(t)e^F=a_l(t)+a_l(-t)+\frac{1}{2t}(e^{\frac{\pi}{n}t}-1)
-\delta_{l,L}\frac{e^{i\mu t+\frac{\pi}{n}Lt}}{t}
-\delta_{l,M}\frac{e^{-i\mu t+\frac{\pi}{n}(M-2L)t}}{t},
\end{eqnarray} 
and
\begin{eqnarray}
G_j(\alpha)=
\left\{\begin{array}{cc}
-ie^{\gamma} \times \alpha,& j\neq L,M,\\
e^{2\gamma} \times \alpha 
\left(-\mu+\frac{\pi i}{n} L-\alpha
\right),& j=L,\\
e^{2\gamma} \times \alpha
\left(\mu+\frac{\pi i}{n}(M-2L)-\alpha
\right)
,& j=M.
\end{array}
\right.
\end{eqnarray}
\begin{eqnarray}
H_0(\beta)=Const.
(\mu+\beta)^{\delta_{L,0}}
\frac{
\Gamma\left(\frac{\beta+\mu}{2\pi i}\right)
\Gamma\left(\frac{\beta-\mu}{2\pi i}+\frac{L}{n}\right)
\Gamma\left(\frac{\beta}{2\pi i}+\frac{1}{2}-\frac{1}{2n}
\right)
}{
\Gamma\left(\frac{\beta+\mu}{2\pi i}+1-\frac{L}{n}\right)
\Gamma\left(\frac{\beta-\mu}{2\pi i}+1+\frac{L-M}{n}\right)
\Gamma\left(\frac{\beta}{2\pi i}\right)}
.\nonumber\\
\end{eqnarray}

We shall prove the characterizing relations
(\ref{character}),
under the specification of the coefficients
$\alpha_{j,k}(t)$ and $\beta_j(t)$.
In what follows we use the auxiliary function,
\begin{eqnarray}
\Delta(\alpha_1,\alpha_2)=
\left(\alpha_1+\alpha_2+\frac{\pi i}{n}\right)
\left(-\alpha_1+\alpha_2+\frac{\pi i}{n}\right)
\left(\alpha_1-\alpha_2+\frac{\pi i}{n}\right)
\left(-\alpha_1-\alpha_2+\frac{\pi i}{n}\right),
\nonumber\\
\end{eqnarray}
which
is invariant under the change of variables $
\alpha_j\to -\alpha_j, (j=1,2)$,
\begin{eqnarray}
\Delta(\alpha_1,\alpha_2)=
\Delta(-\alpha_1,\alpha_2)=
\Delta(\alpha_1,-\alpha_2)=
\Delta(-\alpha_1,-\alpha_2).
\end{eqnarray}
The actions of the operators $Z_j^*(\beta)$
on the boundary state $|B\rangle$ are given by
\begin{eqnarray}
Z_j^*(\beta)|B\rangle
&=&
e^{-2j\gamma}
H_0(\beta) \prod_{l=1}^j \int_{-\infty}^\infty d\alpha_l
\prod_{l=1}^j G_l(\alpha_l)
\prod_{l=1}^j\left(\alpha_l+\alpha_{l-1}+\frac{\pi i}{n}\right)
\nonumber\\
&\times&
\prod_{l=1}^j
\frac{1}{\Delta(\alpha_l,\alpha_{l-1})}
\prod_{l=0}^j
V_l^c(\alpha_l)V_l^c(-\alpha_l)|B\rangle.
\end{eqnarray}
The $0$-th components of the characterizing relation
(\ref{character}) become
\begin{eqnarray}
(LHS)&=&
K(\beta)_0^0 H_0(\beta) V_0^c(\beta)V_0^c(-\beta)|B\rangle,
\\
(RHS)&=&H_0(-\beta)
V_0^c(\beta)V_0^c(-\beta)|B\rangle.
\end{eqnarray}
The $0$-th component of the equation 
(\ref{character})
follows from
\begin{eqnarray}
K(\beta)_0^0=\frac{H_0(-\beta)}{
H_0(\beta)}.
\end{eqnarray}
Next we consider the $0<j<L$ th component
of the equation
(\ref{character}).
The diffence of the both hand side is given by
\begin{eqnarray}
&&K(\beta)_j^jZ_j^*(\beta)|B\rangle-
Z_j^*(-\beta)|B\rangle
\nonumber\\
&=&2e^{-2j\gamma}
\beta H_0(-\beta)
\prod_{l=1}^j \int_{-\infty}^\infty d\alpha_l
\prod_{l=1}^j G_l(\alpha_l)
\prod_{l=2}^j\left(\alpha_l+\alpha_{l-1}+\frac{\pi i}{n}\right)
\nonumber\\
&\times&
\prod_{l=1}^j
\frac{1}{\Delta(\alpha_l,\alpha_{l-1})}
\prod_{l=0}^j
V_l^c(\alpha_l)
V_l^c(-\alpha_l)|B\rangle.
\end{eqnarray}
The following part :
\begin{eqnarray}
\prod_{l=1}^j
\frac{1}{\Delta(\alpha_l,\alpha_{l-1})}
\prod_{l=0}^j
V_l^c(\alpha_l)
V_l^c(-\alpha_l)|B\rangle,
\end{eqnarray}
is invariant under the change of variable
$\alpha_l \leftrightarrow -\alpha_l$.
Therefore a sufficient condition
of the equation,
\begin{eqnarray}
K(\beta)_j^jZ_j^*(\beta)|B\rangle=
Z_j^*(-\beta)|B\rangle,~~(0< j <L),
\end{eqnarray}
is a polynomial identitity,
\begin{eqnarray}
\sum_{\epsilon_1 \cdots \epsilon_j=\pm}
\prod_{l=1}^j \epsilon_l 
\prod_{l=2}^j
\left(\epsilon_l \alpha_l+
\epsilon_{l-1} \alpha_{l-1}+\frac{\pi i}{n}\right)=0,~~
(0< j <L).\label{P1}
\end{eqnarray}
The above equation (\ref{P1}) follows from 
following inductive
relation.
\begin{eqnarray}
&&\sum_{\epsilon_1 \cdots \epsilon_j=\pm}
\prod_{l=1}^j \epsilon_l 
\prod_{l=2}^j
\left(\epsilon_l \alpha_l+
\epsilon_{l-1} \alpha_{l-1}+\frac{\pi i}{n}\right)\nonumber\\
&=&2 \alpha_1 
\sum_{\epsilon_2 \cdots \epsilon_j=\pm}
\prod_{l=2}^j \epsilon_l 
\prod_{l=3}^j
\left(\epsilon_l \alpha_l+
\epsilon_{l-1} \alpha_{l-1}+\frac{\pi i}{n}\right).
\end{eqnarray}
As the same arguments as the above
the characterizing relation (\ref{character})
for $L\leq j<M$,
is reduced to the following polynomial identity,
\begin{eqnarray}
\sum_{\epsilon_1 \cdots \epsilon_j=\pm}
\left(\mu-\frac{\pi i}{n}-
\epsilon_1\alpha_1\right)
\left(\mu-\frac{\pi i}{n} L+\epsilon_L \alpha_L\right)
\prod_{l=1}^j \epsilon_l 
\prod_{l=2}^j
\left(\epsilon_l \alpha_l+
\epsilon_{l-1} \alpha_{l-1}+\frac{\pi i}{n}\right)=0,
\nonumber\\
(L\leq j <M).~~~~~
\label{P2}
\end{eqnarray}
The above equation (\ref{P2}) is 
deformed to the following.
\begin{eqnarray}
&&2^{L} \prod_{l=1}^L \alpha_l
\left(
\mu-\frac{\pi i}{n}L-\alpha_L\right)
\left(
\mu-\frac{\pi i}{n}L+\alpha_L\right)\nonumber\\
&\times&
\sum_{\epsilon_{L+1} \cdots \epsilon_j=\pm}
\prod_{l=L+1}^j \epsilon_l 
\prod_{l=L+2}^j \left(\epsilon_l \alpha_l
+\epsilon_{l-1}\alpha_{l-1}+\frac{\pi i}{n}\right).
\end{eqnarray}
Therefore the equation (\ref{P2}) is reduced to
the equation (\ref{P1}).\\
As the same arguments as the above the characterizing
relation (\ref{character})
for $M\leq j \leq n-1$,
is reduced to the following polynomial identity.
\begin{eqnarray}
&&\sum_{\epsilon_1 \cdots \epsilon_j=\pm}
\left(\mu-\frac{\pi i}{n}L+\epsilon_L \alpha_L \right)
\left(\mu-\frac{\pi i}{n}(2L-M)-\epsilon_M \alpha_M \right)
\nonumber\\
&\times&\prod_{l=1}^j \epsilon_l 
\prod_{l=2}^j
\left(\epsilon_l \alpha_l+
\epsilon_{l-1} \alpha_{l-1}+\frac{\pi i}{n}\right)=0,~~
(M\leq j \leq n-1).\nonumber\\
\label{P3}
\end{eqnarray}
The above equation (\ref{P3})
is reduced to the following.
\begin{eqnarray}
&&2^M \prod_{l=1}^M \alpha_l
\left(\mu-\frac{\pi i}{n}(2L-M)-\alpha_M \right)
\left(\mu-\frac{\pi i}{n}(2L-M)+\alpha_M \right)
\nonumber\\
&\times&
\sum_{\epsilon_{M+1}\cdots \epsilon_j=\pm}
\prod_{l=M+1}^j\epsilon_l
\prod_{l=M+2}^j
\left(\epsilon_l\alpha_l+\epsilon_{l-1}
\alpha_{l-1}+\frac{\pi i}{n}\right).
\end{eqnarray}
Therefore the equation (\ref{P3}) is reduced to
the equation (\ref{P1}).
Now we have proved the characterizing relation
(\ref{character}).

Next we consider the dual boundary state $\langle B|$,
which satisfies
\begin{eqnarray}
K(\beta)_j^j\langle B|Z_j(-\beta+\pi i)=
\langle B|Z_j(\beta+\pi i),~(j=0,\cdots, n-1).
\end{eqnarray}
Here $K(\beta)_j^j$ are diagonal elements of the boundary
$K$-matrix.\\
We make the ansatz that the dual boundary state
has the following form.
\begin{eqnarray}
\langle B|=\langle vac |e^G.
\end{eqnarray}
Here $G$ is a quadratic form of the free bosons,
\begin{eqnarray}
G=\sum_{j,k=1}^{n-1}\int_0^\infty \gamma_{j,k}(t)
a_j(t)a_k(t)dt+
\sum_{j=1}^{n-1}\int_0^\infty \delta_j(t)a_j(t)dt,
\label{def:G}
\end{eqnarray}
where $\gamma_{j,k}(t)$ and $\delta_j(t)$
are scalar functions.\\
As the same arguments as the above
the following coefficients functions 
$\gamma_{j,k}(t)$ and $\delta_j(t)$ of the dual
boundary state are given by
\begin{eqnarray}
\gamma_{j,k}(t)=
-\frac{t 
e^{-\left(\frac{\pi}{n}+2\pi \right)t}}{2}\times I_{j,k}(t),
\label{def:gamma}
\end{eqnarray}
and
\begin{eqnarray}
\delta_j(t)&=&\frac{1}{2}(e^{-\frac{\pi}{n}t}-1)e^{-\pi t}
\sum_{l=1}^{n-1}I_{l,j}(t)\nonumber\\
&-&
e^{i(\mu+\frac{\pi i}{n}(2M+1-L))t}I_{L,j}(t)-
e^{i(-\mu+\frac{\pi i}{n}(2n-M+1))t}I_{M,j}(t).
\label{def:delta}
\end{eqnarray}
The effect of the operator $e^G$ is given by
\begin{eqnarray}
e^G a_l(t) e^{-G}&=&0,~(t>0),\\
e^G a_l(-t) e^{-G}&=&a_l(-t)+e^{-2\pi t}a_l(t)+
\frac{e^{-\pi t}}{t}(e^{\frac{\pi}{n}t}-1)\nonumber\\
&+&
\delta_{L,l}\frac{e^{i\mu t+\frac{\pi}{n}(L-2M)t}}{t}
+\delta_{M,l}\frac{e^{-i\mu t+\frac{\pi}{n}M t-2\pi t}}{t},
~(t>0).
\end{eqnarray}
The following properties are useful.
\begin{eqnarray}
\langle B|V_j^c(\alpha+\pi i)&=&
G_j^*(\alpha)
\langle B|V_j^a(-\alpha+\pi i),~(j=1,\cdots,n-1),\\
\langle B|V_{n}^c(\beta+\pi i)&=&
H_{n}^*(\beta)
\langle B|V_{n}^a(-\beta+\pi i).
\end{eqnarray}
Here we have set
\begin{eqnarray}
G_j^*(\alpha)=
\left\{\begin{array}{cc}
ie^\gamma \times \alpha,& j\neq L,M,\\
\alpha \times
\left(-\mu+\frac{\pi i}{n}(n+L-2M)+\alpha
\right)^{-1},& j=L,\\
\alpha \times 
\left(
\mu-\frac{\pi i}{n}(n-M)+\alpha \right)^{-1},& j=M,
\end{array}\right.
\end{eqnarray}
and

\begin{eqnarray}
H_{n}^*(\beta)=Const.
(\mu-\beta)^{\delta_{M,n}}
\frac{
\Gamma\left(\frac{-\beta+\mu}{2\pi i}\right)
\Gamma\left(\frac{-\beta-\mu}{2\pi i}+\frac{L}{n}\right)
\Gamma\left(\frac{-\beta}{2\pi i}+\frac{1}{2}-\frac{1}{2n}
\right)
}{
\Gamma\left(\frac{-\beta+\mu}{2\pi i}+1-\frac{L}{n}\right)
\Gamma\left(\frac{-\beta-\mu}{2\pi i}+1+\frac{L-M}{n}\right)
\Gamma\left(\frac{-\beta}{2\pi i}\right)}
.\nonumber\\
\end{eqnarray}
The function $H_n^*(\beta)$ satisfies
\begin{eqnarray}
k_{L,M}(\beta)
=
\frac{H_n^*(\beta)}{H_n^*(-\beta)}
\left(\frac{\mu+\beta}{\mu-\beta}\right)^{\delta_{M,n}}.
\end{eqnarray}

~\\
{\it
Let us summarize the results of this section.
We have constructed
the free field realization of the boundary state $|B\rangle$
and the dual boundary state $\langle B|$.}
\begin{eqnarray}
\langle B|=\langle vac |e^{G},~~
e^F| vac \rangle=|B\rangle.
\end{eqnarray}
{\it Here $F$ and $G$ are quadratic forms of free bosons.
The explicit formulae of $F$ and $G$
are given by (\ref{def:F}), (\ref{def:G}).
The coefficients $\alpha_{j,k}(t)$
,$\beta_j(t)$, $\delta_{j,k}(t)$,
$\gamma_{j}(t)$, and $I_{j,k}(t)$ are given 
by (\ref{def:alpha}), (\ref{def:beta}),
(\ref{def:gamma}), (\ref{def:delta}) and (\ref{def:Inverse}),
respectively.}

\section{Form Factors}
The purpose of this section is to give
integral representations of the form factors of the local
fields,
defined by
\begin{eqnarray}
&&f(\delta_1,\cdots,\delta_M|
\beta_1,\cdots,\beta_N)_{j_1,\cdots,j_N}^{k_1,\cdots,k_M}
\nonumber\\
&=&
\frac{1}{\langle B|B \rangle}\times
\langle B|Z_{k_1}'(\delta_1)
\cdots Z_{k_M}'(\delta_M)
|\beta_1,\cdots,\beta_N\rangle_{j_1,
\cdots, j_N}.\label{def:form}
\end{eqnarray}
Here $\langle B|$ and $|B\rangle$ are the boundary state
and it's dual.
The state $|\beta_1,\cdots,\beta_N\rangle
_{j_1,\cdots,j_N}$ is the basis of the space of the physical
states defined in the section 2.
The operators $Z_{k}'(\delta), (k=0,\cdots, n-1)$
are the local operators given in the next section.

\subsection{Local Operators}

In terminology of the Quantum Field Thory,
the local operator is the one which commutes
with the Zamolodchikov-Faddeev operators, up to
scalar function multiplicity.
In this subsection
we give the free field realization of a class of the local
operators $Z_j'(\beta)$ of the present model.
Naively the local operators of the present model
are a limiting case of the type-I vertex operators
$\Phi_j(\beta)$ of the $A_{n-1}^{(1)}$ critical chain
\cite{KY}.

In this section,
we give the free field realization
of the local operators $Z_j'(\delta)$
which satisfy the following commutation relations.
\begin{eqnarray}
Z_j^*(\beta_1)Z_k'(\beta_2)=Const.
{\cal L}(\beta_1-\beta_2)
Z_k'(\beta_2)Z_j^*(\beta_1), \label{Com3}
\end{eqnarray}
where we set
\begin{eqnarray}
{\cal L}(\beta)=\frac{\Gamma\left(\frac{\beta}{2\pi i}
+1-\frac{1}{2n}\right)
\Gamma\left(\frac{-\beta}{2\pi i}+\frac{1}{2n}\right)
}{
\Gamma\left(\frac{-\beta}{2\pi i}
+1-\frac{1}{2n}\right)
\Gamma\left(\frac{\beta}{2\pi i}+\frac{1}{2n}\right)
}.
\end{eqnarray}
Let us set the auxiliary fields $b_j(t), (1\leq j \leq n-1,
t \in {\mathbb{R}})$ by
\begin{eqnarray}
b_j(t)=e^{-\frac{\pi}{n}|t|}\times a_j(t).
\end{eqnarray}
The bose field $b(t)$ satisfies
the following commutation relation.
\begin{eqnarray}
[b_j(t),b_k(t')]=-\frac{1}{t}
\frac{{\rm sh}\frac{(a_j|a_k)\pi t}{n}}{{\rm sh}\frac{\pi t}{n}}
e^{-\frac{\pi}{n}|t|}\delta(t+t').
\end{eqnarray}
Here $((a_j|a_k))_{1\leq j,k \leq n-1}$
is the Cartan matrix of type $A_{n-1}$.\\
\begin{eqnarray}
b_1^*(t)=-\sum_{j=1}^{n-1}b_j(t)
\frac{{\rm sh}\frac{(n-j)\pi t}{n}}{
{\rm sh}\pi t}.
\end{eqnarray}
We have
\begin{eqnarray}
[b_1^*(t),b_j(t')]=\delta_{j,1}
\frac{e^{-\frac{\pi}{n}|t|}}{t}\delta(t+t').
\end{eqnarray}
Let us introduce the basic operators,
\begin{eqnarray}
U_j(\delta)&=&:\exp\left(
-\int_{-\infty}^\infty
b_j(t)e^{i\delta t}dt\right): (1\leq j \leq n-1),\\
U_0(\delta)&=&:
\exp\left(-\int_{-\infty}^\infty b_1^*(t)
e^{i \delta t}dt\right):.
\end{eqnarray}
The basic operators satisfy the following contraction relations.
\begin{eqnarray}
U_j(\delta_1)U_{j-1}(\delta_2)=
\frac{-ie^{-\gamma}}
{\delta_2-\delta_1-\frac{\pi i}{n}}
:U_j(\delta_1)U_{j-1}(\delta_2):,~(j=1,\cdots,n-1),\\
U_{j-1}(\delta_1)U_{j}(\delta_2)=
\frac{-ie^{-\gamma}}
{\delta_2-\delta_1-\frac{\pi i}{n}}
:U_{j-1}(\delta_1)U_{j}(\delta_2):,~(j=1,\cdots,n-1),\\
U_j(\delta_1)U_j(\delta_2)=
e^{2\gamma}(\delta_1-\delta_2)
\left(\delta_2-\delta_1-\frac{2\pi i}{n}\right)
:U_j(\delta_1)U_j(\delta_2):,(j=1,\cdots,n-1),
\end{eqnarray}
\begin{eqnarray}
U_0(\delta_1)U_0(\delta_2)=e^{(\gamma+{\rm log}2\pi)\frac{n-1}{n}}
\frac{\Gamma\left(\frac{\delta_1-\delta_2}{2\pi i}+1\right)}
{\Gamma\left(\frac{\delta_1-\delta_2}{2\pi i}+\frac{1}{n}\right)}
:U_0(\delta_1)U_0(\delta_2):
\end{eqnarray}

~\\
We give the free field realization of the local
operators $Z_j'(\delta), (j=1,\cdots,n-1)$.
\begin{eqnarray}
Z_j'(\delta)&=&\int_{-\infty}^\infty d\gamma_1
\cdots \int_{-\infty}^\infty
d\gamma_j
\frac{U_0(\delta)U_1(\gamma_1)\cdots U_j(\gamma_j)}{
\prod_{k=1}^j
\left(\gamma_{k-1}-\gamma_k-\frac{\pi i}{n}\right)},\nonumber\\
&=&(-ie^{-\gamma})^j \int_{-\infty}^\infty d\gamma_1
\cdots \int_{-\infty}^\infty
d\gamma_j\nonumber\\
&\times&
:U_0(\delta)U_1(\gamma_1)\cdots U_j(\gamma_j)
:\prod_{k=1}^j
\frac{1}{\left(
\gamma_{k-1}-\gamma_k-\frac{\pi i}{n}
\right)\left(\gamma_k-\gamma_{k-1}-\frac{\pi i}{n}\right)},
\end{eqnarray}
where we set $\delta=\gamma_0$.
The local operators $Z_j'(\delta)$ satisfy the
commutation relations similar to (\ref{Com1}).
See the refernce \cite{MT}.
The two type basic operators $U_j(\delta)$ and $V_k(\beta)$
satisfy the following interaction relations.
\begin{eqnarray}
V_j(\beta_1)U_{j-1}(\beta_2)=
\frac{ie^{-\gamma}}{\beta_1-\beta_2}
:V_j(\beta_1)U_{j-1}(\beta_2):,(j=1,\cdots,n-1),\\
V_{j-1}(\beta_1)U_j(\beta_2)=
\frac{ie^{-\gamma}}{\beta_1-\beta_2}
:V_{j-1}(\beta_1)U_j(\beta_2):,(j=1,\cdots,n-1),
\end{eqnarray}
\begin{eqnarray}
V_j(\beta_1)U_j(\beta_2)=
-e^{2\gamma}\left(\beta_1-\beta_2+\frac{\pi i}{n}\right)
\left(\beta_2-\beta_1+\frac{\pi i}{n}\right)
:V_j(\beta_1)U_j(\beta_2):,\\
U_j(\beta_1)V_j(\beta_2)=
-e^{2\gamma}\left(\beta_1-\beta_2+\frac{\pi i}{n}\right)
\left(\beta_2-\beta_1+\frac{\pi i}{n}\right)
:U_j(\beta_1)V_j(\beta_2):,\\
(j=1,\cdots,n-1),\nonumber
\end{eqnarray}
\begin{eqnarray}
V_0(\beta_1)U_0(\beta_2)=
e^{(\gamma+{\rm log}2\pi)\frac{n-1}{n}}
\frac{\Gamma\left(\frac{\beta_1-\beta_2}{2\pi i}
+1-\frac{1}{2n}\right)
}{
\Gamma\left(\frac{\beta_1-\beta_2}{2\pi i}+\frac{1}{2n}\right)
}
:V_0(\beta_1)U_0(\beta_2):.
\end{eqnarray}
The commutation relation (\ref{Com3}) is proved as the
same manner as the reference \cite{JKM}.

\subsection{Integral Representations}

In this section we calculate explicit formulae
of the matrix elements,
\begin{eqnarray}
f(\delta_1, \cdots ,\delta_M
|\beta_1 \cdots \beta_N)
_{j_1 \cdots j_N}^{k_1 \cdots k_M}.
\end{eqnarray}
In order to evaluate the above expectation value,
we invoke the free field realization of
the Zamolodchikov-Faddeev operaors, the local operators,
and the boundary state and it's dual state.

Fix the indexes
$\{j_1,\cdots,j_N\}$, where $j_1,\cdots, j_N \in \{
0,1, \cdots,n-1\}$, and $\{k_1,\cdots,k_M\}
$, where $
k_1,\cdots, k_M \in \{
0,1, \cdots,n-1\}
$.
We associate the integration variables 
$\alpha_{j,r},~(1\leq r \leq N, 1\leq j \leq j_r)$
to the basic operator $V_j(\alpha_{j,r})$ contained in
the Zamolodchikov-Faddeev operator $Z_{j_r}^*(\beta_r)$.
We also use the notation $\alpha_{0,r}=\beta_r$.
We associate the integration variables
$\gamma_{k,s},~(1\leq s \leq N, 1\leq k \leq k_s)$
to the basic operator $U_j(\gamma_{k,s})$ contained
in the local operator $Z_{k_s}'(\delta_s)$.
We also use the notation $\gamma_{0,s}=\delta_s$.
Let us set the index set
${\cal A}_j$ and ${\cal G}_k$ by 
\begin{eqnarray}
{\cal A}_j=\{r|j_r \geq j\},~~
{\cal G}_k=\{s|k_s \geq k\}.
\end{eqnarray}
By normal-ordering 
the prodct of the Zamolodchikov-Faddeev operators and the
local operators,
we have the following formula.
\begin{eqnarray}
&&f(\delta_1, \cdots ,\delta_M
|\beta_1 \cdots \beta_N)
_{j_1 \cdots j_N}^{k_1 \cdots k_M}\nonumber\\
&=&
E(\{\beta\}|\{\delta\})
\prod_{r=1}^N \prod_{j=1}^{j_r}\int_{-\infty}^\infty
d\alpha_{j,r}
\prod_{s=1}^M \prod_{k=1}^{k_s} \int_{-\infty}^\infty
d\gamma_{k,s}
I(\{\alpha\}|\{\gamma\})_{j_1 \cdots j_N}
^{k_1 \cdots k_M}.\label{def:integral}
\end{eqnarray}
Here we set $E(\{\beta\}|\{\delta\})$ by
\begin{eqnarray}
&&E(\{\beta\}|\{\delta\})
=
\prod_{r=1}^N\prod_{s=1}^M
\frac{\Gamma\left(
\frac{\beta_r-\delta_s}{2\pi i}+1-\frac{1}{2n}
\right)}{
\Gamma\left(
\frac{\beta_r-\delta_s}{2\pi i}+\frac{1}{2n}\right)
}
\nonumber\\
&\times&
\prod_{1\leq r_1<r_2 \leq N}
\frac{\Gamma\left(
\frac{\beta_{r_1}-\beta_{r_2}}{2\pi i}+1-\frac{1}{n}\right)}
{\Gamma\left(\frac{\beta_{r_1}-\beta_{r_2}}{2\pi i}\right)}
\prod_{1\leq s_1<s_2 \leq M}
\frac{
\Gamma\left(
\frac{\delta_{s_1}-\delta_{s_2}}{2\pi i}+1\right)
}{
\Gamma\left(\frac{\delta_{s_1}-\delta_{s_2}}{2\pi i}
+\frac{1}{n}\right)
}.\label{def:E}
\end{eqnarray}
Here we set the integrand function by
\begin{eqnarray}
&&I(\{\alpha\}|\{\gamma\})_{j_1 \cdots j_N}
^{k_1 \cdots k_M}\nonumber
\\
&=&
\prod_{j=1}^{n-1}
\left\{
\prod_{r_1,r_2 \in {\cal A}_j
\atop{r_1<r_2}
}\left(\alpha_{j,r_1}-\alpha_{j,r_2}\right)
\left(\alpha_{j,r_1}-\alpha_{j,r_2}
-\frac{2\pi i}{n}\right)
\right\}\nonumber\\
&\times&
\prod_{j=1}^{n-1}
\left\{
\prod_{r_1 \in {\cal A}_j,
r_2 \in {\cal A}_{j-1}
\atop{r_1 \leq r_2}}
\left(\alpha_{j-1,r_1}-\alpha_{j,r_2}
-\frac{\pi i}{n}\right)^{-1}
\prod_{r_1 \in {\cal A}_{j-1},
r_2 \in {\cal A}_{j}
\atop{r_1 \leq r_2}}
\left(\alpha_{j,r_1}-\alpha_{j-1,r_2}
-\frac{\pi i}{n}\right)^{-1}
\right\}
\nonumber\\
&\times&\prod_{k=1}^{n-1}
\left\{
\prod_{s_1,s_2 \in {\cal G}_j
\atop{s_1<s_2}
}\left(\gamma_{k,s_1}-\gamma_{k,s_2}\right)
\left(\gamma_{k,s_1}-\gamma_{k,s_2}
+\frac{2\pi i}{n}\right)
\right\}\nonumber\\
&\times&
\prod_{k=1}^{n-1}
\left\{
\prod_{s_1 \in {\cal G}_j,
s_2 \in {\cal G}_{j-1}
\atop{s_1 \leq s_2}}
\left(\gamma_{k-1,s_1}-\gamma_{k,s_2}
+\frac{\pi i}{n}\right)^{-1}
\prod_{s_1 \in {\cal G}_{j-1},
s_2 \in {\cal G}_{j}
\atop{s_1 \leq s_2}}
\left(\gamma_{k,s_1}-\gamma_{k-1,s_2}
+\frac{\pi i}{n}\right)^{-1}
\right\}\nonumber\\
&\times&
\prod_{j=1}^{n-1}
\left\{
\prod_{s \in {\cal G}_j,
r \in {\cal A}_j}\left(
\alpha_{j,r}-\gamma_{j,s}+\frac{\pi i}{n}
\right)\left(
\gamma_{j,s}-\alpha_{j,r}+\frac{\pi i}{n}
\right)\right\}\nonumber\\
&\times&
\prod_{j=1}^{n-1}
\left\{
\prod_{s \in {\cal G}_j}
\prod_{r \in {\cal A}_{j-1}}
\left(\gamma_{j,s}-\alpha_{j-1,r}\right)^{-1}
\prod_{s \in {\cal G}_{j-1}}
\prod_{r \in {\cal A}_{j}}
\left(\gamma_{j-1,s}-\alpha_{j,r}\right)^{-1}
\right\}\times
J(\{\alpha\}|\{\gamma\})_{j_1 \cdots j_N}
^{k_1 \cdots k_M}.\nonumber\\
\label{def:I}
\end{eqnarray}
Here we have set 
%\begin{eqnarray}
%&&J(\{ \alpha \}|
%\{\gamma \})_{j_1 \cdots j_N}
%^{k_1 \cdots k_M}
%=\frac{1}{\langle B|B\rangle}
%\times
%\langle B|
%\exp\left(
%\int_0^\infty
%\sum_{r=1}^N
%\left\{
%a_1^*(-t)e^{-i 
%\beta_r t}+\sum_{j=1}^{j_r}
%a_j(-t)e^{-i \alpha_{j,r}t}\right\}
%dt
%\right.\nonumber\\
%&&\left.~~~~~~~~~~~~~~~-
%\int_0^\infty
%\sum_{s=1}^{M}
%\left\{
%b_1^*(-t)e^{-i \delta_s t}
%+\sum_{k=1}^{k_s}
%b_j(-t)e^{-i\gamma_{k,s}t}
%\right\}dt \right)\nonumber\\
%&\times&\exp\left(
%\int_0^\infty
%\sum_{r=1}^N
%\left\{
%a_1^*(t)e^{i 
%\beta_{r}t}+
%\sum_{j=1}^{j_r}
%a_j(t)e^{i\alpha_{j,r}
%t}
%\right\}dt\right.
%\nonumber\\
%&&~~~~~~~~~~~~~~~~~~-\left.
%\sum_{s=1}^{M}\int_{-\infty}^\infty
%\left\{
%b_1^*(t)e^{i \gamma_{s}t}
%+
%\sum_{k=1}^{k_s}
%b_k(t)e^{i \delta_{k,s} t}
%\right\}dt\right)
%|B\rangle.
%\end{eqnarray}
\begin{eqnarray}
&&J(\{\alpha\}|\{\gamma\})_{j_1 \cdots j_N}^{k_1 \cdots k_M}
\nonumber\\
&=&
\frac{1}{\langle B|B\rangle}
\times
\langle B|
\exp\left(\int_0^\infty
X_j(t)a_j(-t)
dt\right)
\exp\left(\int_0^\infty
Y_j(t)a_j(t)
dt\right)
|B\rangle,
\end{eqnarray}
where
\begin{eqnarray}
X_j(t)=\frac{{\rm sh}\frac{(n-j)\pi t}{n}}{{\rm sh}\pi t}
\left(-\sum_{r=1}^N e^{-i\beta_r t}+
e^{-\frac{\pi}{n}t}
\sum_{s=1}^M e^{-i\delta_s t}\right)
+\sum_{r \in {\cal A}_j}e^{-i \alpha_{j,r}t}-
e^{-\frac{\pi}{n}t}
\sum_{s \in {\cal G}_j}e^{-i\gamma_{j,s} t},
\\
Y_j(t)=
\frac{{\rm sh}\frac{(n-j)\pi t}{n}}{{\rm sh}\pi t}
\left(-\sum_{r=1}^N e^{i\beta_r t}+
e^{-\frac{\pi}{n}t}
\sum_{s=1}^M e^{i\delta_s t}\right)
+\sum_{r \in {\cal A}_j}e^{i \alpha_{j,r}t}-
e^{-\frac{\pi}{n}t}
\sum_{s \in {\cal G}_j}e^{i\gamma_{j,s} t}.
\end{eqnarray}

Next we evaluate the 
vacuum expectation value,
$
J(\{\alpha\}|\{\gamma\})_{j_1 \cdots j_N}
^{k_1 \cdots k_M}
$.
In what follows we use the abberiviation,
\begin{eqnarray}
A_{j,k}(t)=-\frac{1}{t}\frac{
{\rm sh}\frac{(a_j|a_k)\pi t}{n}}{
{\rm sh}\frac{\pi t}{n}}e^{\frac{\pi}{n}|t|},~~(
j,k=0,\cdots,n-1).
\end{eqnarray}

For our purpose we use the coherent states,
$
|\xi_1,\cdots,\xi_{n-1}\rangle
$,
defined by
\begin{eqnarray}
|\xi_1,\cdots,\xi_{n-1}\rangle=
\exp\left(
\sum_{k=1}^{n-1}\int_0^\infty
\xi_k(s) a_k(-s)ds \right)|vac \rangle,
\end{eqnarray}
and it's dual states,$
\langle \bar{\xi}_1,\cdots,\bar{\xi}_{n-1}|
$, 
\begin{eqnarray}
\langle \bar{\xi}_1,\cdots,\bar{\xi}_{n-1}|=
\langle vac|\exp\left(
\sum_{k=1}^{n-1}\int_0^\infty
\bar{\xi}_k(s)a_k(s)ds \right).
\end{eqnarray}
The coherent states enjoy
\begin{eqnarray}
a_j(t)|\xi_1,\cdots,\xi_{n-1}\rangle=
\sum_{k=1}^{n-1}
A_{j,k}(t)
\xi_k(t)|\xi_1,\cdots,\xi_{n-1}\rangle,~~~(t>0),\\
\langle \bar{\xi}_1,\cdots,\bar{\xi}_{n-1}|a_j(-t)
=\sum_{k=1}^{n-1}
A_{j,k}(t)
\bar{\xi}_k(t)
\langle \bar{\xi}_1,\cdots,\bar{\xi}_{n-1}|
,~~~(t>0).
\end{eqnarray}
For our purpose the following completeness relation
by means
of Feynmann path integral
is useful.
\begin{eqnarray}
id&=&Const.\times
\int
\prod_{k=1}^{n-1}
\prod_{s>0}
d\xi_k(s) d\bar{\xi}_k(s)\nonumber\\
&\times&
\exp\left(-\sum_{k_1,k_2=1}^{n-1}
\int_0^\infty
A_{k_1,k_2}(s)
\xi_{k_1}(s)
\bar{\xi}_{k_2}(s) ds\right)
|\xi_1,\cdots,\xi_{n-1}\rangle
\langle \bar{\xi}_1,\cdots,\bar{\xi}_{n-1}|.\nonumber\\
\end{eqnarray}
Here the integration $\int d\xi d\bar{\xi}$ is taken over
the entire complex plane with the measure
$d\xi d\bar{\xi}=-2i dx dy$ for $\xi=x+iy$.
\\
In what follows we use the following
abberiviations.
\begin{eqnarray}
\widetilde{\beta_j}(t)&=&\sum_{k=1}^{n-1}
A_{j,k}(t)\beta_k(t)=
\frac{1}{2t}(e^{\frac{\pi}{n}t}-1)-\frac{
e^{i\mu t+\frac{\pi}{n}Lt}}{t}\delta_{j,L}
-\frac{e^{-i\mu t+\frac{\pi}{n}(M-2L)t}}{t}
\delta_{j,M},\\
\widetilde{\delta_j}(t)&=&\sum_{k=1}^{n-1}
A_{j,k}(t)\delta_k(t)=
\frac{e^{-\pi t}}{2t}(e^{\frac{\pi}{n}t}-1)
+\frac{e^{i\mu t+\frac{\pi}{n}(L-2M)t}}{t}
\delta_{j,L}+
\frac{e^{-i\mu t +\frac{\pi}{n}(M-2n)t}}{t}
\delta_{j,M},
\nonumber\\
\end{eqnarray}
\begin{eqnarray}
\widetilde{X}_j(t)&=&\sum_{k=1}^{n-1}A_{j,k}(t)
X_k(t)\\
&=&
\frac{e^{\frac{\pi}{n}t}}{t}
\left(\sum_{r \in {\cal A}_{j-1}}e^{-i\alpha_{j-1,r}t}+
\sum_{r \in {\cal A}_{j+1}}e^{-i\alpha_{j+1,r}t}\right)
-\frac{1}{t}(1+e^{\frac{2 \pi}{n}t})
\sum_{r \in {\cal A}_j}e^{-i\alpha_{j,r}t}\nonumber
\\
&-&
\frac{1}{t}
\left(\sum_{s \in {\cal G}_{j-1}}e^{-i\gamma_{j-1,s}t}+
\sum_{s \in {\cal G}_{j+1}}e^{-i\gamma_{j+1,s}t}\right)
+\frac{e^{-\frac{\pi}{n}t}}{t}(1+e^{\frac{2 \pi}{n}t})
\sum_{s \in {\cal G}_j}e^{-i\gamma_{j,s}t},\nonumber
\\
\widetilde{Y}_j(t)&=&\sum_{k=1}^{n-1}
A_{j,k}(t)Y_k(t)=\widetilde{X}_j(t)^*.
\end{eqnarray}
Here we understand ${\cal A}_{n}={\cal G}_n=\{\emptyset\}$.\\
Using the Bogoliubov transformation of $e^F$,
we have
\begin{eqnarray}
&&J(\{ \alpha \}|
\{\gamma \})_{j_1 \cdots j_N}
^{k_1 \cdots k_M}\nonumber\\
&=&\frac{1}{\langle vac |e^G e^F |vac \rangle}
\times\langle vac| e^G e^F \exp\left(
\int_0^\infty
\sum_{j=1}^{n-1}(X_j(t)+Y_j(t))a_j(-t)dt\right)|vac \rangle
\nonumber
\\
&\times&
\exp\left(\int_0^\infty
\sum_{j=1}^{n-1}\widetilde{\beta}_j(t)Y_j(t)dt
\right)
\exp\left(
\frac{1}{2}\int_0^\infty
\sum_{j_1,j_2=0}^{n-1}
\widetilde{Y}_{j_1}(t)Y_{j_2}(t)dt
\right).\nonumber\\
\end{eqnarray}
We then insert 
the completeness relation of the coherent states
between $e^G$ and $e^F$,
we have without-operator formula.
\begin{eqnarray}
&&\langle vac| e^G e^F \exp\left(
\int_0^\infty
\sum_{j=1}^{n-1}(X_j(t)+Y_j(t))a_j(-t)dt\right)|vac \rangle
\nonumber\\
&=&
\int \prod_{j=1}^{n-1}\prod_{t>0}
d\xi_j(t)d\bar{\xi}_j(t)
\exp\left(
\int_0^\infty
\sum_{j=1}^{n-1}(\widetilde{\delta}_j(t)\xi_j(t)
+(\widetilde{\beta}_j(t)+
\widetilde{X}_j(t)+\widetilde{Y}_j(t))\bar{\xi}_j(t))dt
\right.\nonumber\\
&+&\left.
\int_0^\infty
\sum_{j_1,j_2=0}^{n-1}
A_{j_1,j_2}(t)
\left(\frac{e^{-2\pi t}}{2}\xi_{j_1}(t)\xi_{j_2}(t)
-\xi_{j_1}(t)\bar{\xi}_{j_2}(t)+
\frac{1}{2}\bar{\xi}_{j_1}(t)\bar{\xi}_{j_2}(t)\right)dt
\right).
\end{eqnarray}
Performing the Gau$\beta$ integral
(quadratic integral) calculations,
\begin{eqnarray}
\int_{-\infty}^\infty e^{-x^2}dx=\sqrt{\pi},
\end{eqnarray}
we get the following formula.
\begin{eqnarray}
&&J(\{ \alpha \}|
\{\gamma \})_{j_1 \cdots j_N}
^{k_1 \cdots k_M}\nonumber\\
&=&\exp\left(
\int_0^\infty
\frac{-te^{-\frac{\pi}{n}t}}{1-e^{-2\pi t}}
\left\{
\sum_{l=1}^{n-1}I_{l,l}(t)
\left(\frac{e^{-2\pi t}}{2}\widetilde{X}_l(t)^2
+e^{-2\pi t}\widetilde{X}_l(t)\widetilde{Y}_l(t)
+\frac{1}{2}\widetilde{Y}_l(t)^2\right)\right.\right.
\nonumber\\
&+&
\sum_{1\leq l_1<l_2 \leq n-1}I_{l_1,l_2}(t)
(e^{-2\pi t}\widetilde{X}_{l_1}(t)\widetilde{Y}_{l_2}(t)
+e^{-2\pi t}\widetilde{X}_{l_2}(t)\widetilde{Y}_{l_1}(t)
+
e^{-2\pi t}\widetilde{X}_{l_1}(t)\widetilde{X}_{l_2}(t)
+\widetilde{Y}_{l_1}(t)\widetilde{Y}_{l_2}(t)
)
\nonumber
\\
&+&
\sum_{l=1}^{n-1}I_{l,l}(t)
\left(\widetilde{\beta}_l(t)
(e^{-2\pi t}\widetilde{X}_l(t)+\widetilde{Y}_l(t))
+\widetilde{\delta}_l(t)(
\widetilde{X}_l(t)+\widetilde{Y}_l(t))
\right)\nonumber\\
&+&
\sum_{1\leq l_1<l_2 \leq n-1}I_{l_1,l_2}(t)
\left(\widetilde{\beta}_{l_1}(t)
(e^{-2\pi t}\widetilde{X}_{l_2}(t)+\widetilde{Y}_{l_2}(t))
+\widetilde{\beta}_{l_2}(t)
(e^{-2\pi t}\widetilde{X}_{l_1}(t)+\widetilde{Y}_{l_1}(t))\right.
\nonumber\\
&+&\left.\left.\left.
\widetilde{\delta}_{l_1}(t)(
\widetilde{X}_{l_2}(t)+\widetilde{Y}_{l_2}(t))
+
\widetilde{\delta}_{l_2}(t)(
\widetilde{X}_{l_1}(t)+\widetilde{Y}_{l_1}(t))
\right)
\right\}dt\right).
\end{eqnarray}
Therefore we can write $
J(\{ \alpha \}|
\{\gamma \})_{j_1 \cdots j_N}
^{k_1 \cdots k_M}
$ by using multi-Gamma functions.
We arrive at the following formula.
\begin{eqnarray}
J(\{ \alpha \}|
\{\gamma \})_{j_1 \cdots j_N}
^{k_1 \cdots k_M}=
J_{B}(\{ \alpha \}|
\{\gamma \})_{j_1 \cdots j_N}
^{k_1 \cdots k_M}
\times
J_{F}(\{ \alpha \}|
\{\gamma \})_{j_1 \cdots j_N}
^{k_1 \cdots k_M},\label{def:J}
\end{eqnarray}
where the factor $J_{B}(\{ \alpha \}|
\{\gamma \})_{j_1 \cdots j_N}
^{k_1 \cdots k_M}$ depends on the boundary 
condition,
and the factor $
J_{F}(\{ \alpha \}|
\{\gamma \})_{j_1 \cdots j_N}
^{k_1 \cdots k_M}
$ is free from the boundary condition.
\begin{eqnarray}
&&J_
{F}(\{ \alpha \}|
\{\gamma \})_{j_1 \cdots j_N}
^{k_1 \cdots k_M}=
\prod_{l=1}^{n-1}
\digamma_{l,l}^{(1)}
(\{\alpha+2\pi\},\{\gamma+2\pi\}
|\{-\alpha\},\{-\gamma\})\nonumber\\
&\times&
\prod_{l=1}^{n-1}
\sqrt{
\digamma_{l,l}^{(1)}(\{\alpha-2\pi i\},\{\gamma-2\pi i\}|
\{\alpha\},\{\gamma\})}
\cdot \sqrt{
\digamma_{l,l}^{(1)}(\{-\alpha\},\{-\gamma\}|
\{-\alpha\},\{-\gamma\})
}\nonumber\\
&\times&
\prod_{l_1,l_2=1 \atop{l_1<l_2}}^{n-1}
\digamma_{l_1,l_2}^{(1)}
(\{\alpha-2\pi i\},\{\gamma-2\pi i\}|\{-\alpha\},\{-\gamma\})
\cdot \digamma_{l_1,l_2}^{(1)}
(\{-\alpha-2\pi i\},\{-\gamma-2\pi i\}|\{\alpha\},\{\gamma\})
\nonumber\\
&\times&
\prod_{l_1,l_2 \atop{l_1<l_2}}^{n-1}
\digamma_{l_1,l_2}^{(1)}
(\{\alpha-2\pi i\},\{\gamma-2\pi i\}|\{\alpha\},\{\gamma\})
\cdot \digamma_{l_1,l_2}^{(1)}
(\{-\alpha\},\{-\gamma\}|\{-\alpha\},\{-\gamma\})
\nonumber\\
&\times&\prod_{l=1}^{n-1}
\digamma_{l,l,l}^{(2)}(\{\alpha-\pi i\},\{\gamma-\pi i\})
\cdot
\digamma_{l,l,l}^{(2)}(\{-\alpha\},\{-\gamma\})\nonumber\\
&\times&
\prod_{l_1,l_2=1
\atop{l_1<l_2}}^{n-1}
\digamma_{l_1,l_2,l_1}^{(2)}(\{\alpha-\pi i\},\{\gamma-\pi i\})
\cdot \digamma_{l_1,l_2,l_2}^{(2)}
(\{\alpha-\pi i\},\{\gamma-\pi i\})
\nonumber\\
&\times&
\prod_{l_1,l_2=1
\atop{l_1<l_2}}^{n-1}
\digamma_{l_1,l_2,l_1}^{(2)}(\{-\alpha\},\{-\gamma\})
\cdot \digamma_{l_1,l_2,l_2}^{(2)}(\{-\alpha\},\{-\gamma\}).
\label{def:JF}
\end{eqnarray}

~\\
Here we have set 
\begin{eqnarray}
&&\digamma_{l_1,l_2}^{(1)}(\{\alpha\},\{\gamma\}|
\{\alpha'\},\{\gamma'\})\nonumber\\
&=&
\prod_{r_1 \in {\cal A}_{l_1-1}}
\frac{
\Omega_{l_1,l_2}^{{\cal A}}\left(
\left.\alpha_{l_1-1,r_1}
+\frac{\pi i}{n}\right|\{\alpha' \}\right)}
{
\Omega_{l_1,l_2}^{{\cal G}}\left(
\left.\alpha_{l_1-1,r_1}
+\frac{\pi i}{n}\right|\{\gamma'-\frac{\pi i}{n} \}\right)
}
\prod_{r_1 \in {\cal A}_{l_1+1}}
\frac{
\Omega_{l_1,l_2}^{{\cal A}}\left(
\left.\alpha_{l_1+1,r_1}
+\frac{\pi i}{n}\right|\{\alpha' \}\right)}
{
\Omega_{l_1,l_2}^{{\cal G}}\left(
\left.\alpha_{l_1+1,r_1}
+\frac{\pi i}{n}\right|\{\gamma'-\frac{\pi i}{n} \}\right)
}\nonumber\\
&\times&
\prod_{r_1 \in {\cal A}_{l_1}}
\frac{
\Omega_{l_1,l_2}^{{\cal G}}
\left(\alpha_{l_1,r_1}\left|
\{\gamma'-\frac{\pi i}{n}\}\right.\right)
\Omega_{l_1,l_2}^{{\cal G}}
\left(\left.\alpha_{l_1,r_1}+\frac{2\pi i}{n}\right|
\{\gamma'-\frac{\pi i}{n}\}\right)
}{
\Omega_{l_1,l_2}^{{\cal A}}(
\alpha_{l_1,r_1}|\{\alpha'\})
\Omega_{l_1,l_2}^{{\cal A}}
\left(\left.\alpha_{l_1,r_1}+\frac{2\pi i}{n}\right|\{\alpha'\} 
\right)
}\nonumber\\
&\times&
\prod_{s_1 \in {\cal G}_{l_1-1}}
\frac{
\Omega_{l_1,l_2}^{{\cal G}}\left(
\left.\gamma_{l_1-1,s_1}\right|
\{\gamma'-\frac{\pi i}{n} \}\right)}
{
\Omega_{l_1,l_2}^{{\cal G}}\left(
\left.\alpha_{l_1-1,s_1}
\right|\{\gamma' \}\right)
}
\prod_{r_1 \in {\cal A}_{l_1+1}}
\frac{
\Omega_{l_1,l_2}^{{\cal G}}\left(
\left.\gamma_{l_1+1,s_1}
\right|\{\gamma' -\frac{\pi i}{n}\}\right)}
{
\Omega_{l_1,l_2}^{{\cal A}}\left(
\left.\gamma_{l_1+1,s_1}
\right|\{\gamma' \}\right)
}\nonumber\\
&\times&
\prod_{s_1 \in {\cal G}_{l_1}}
\frac{
\Omega_{l_1,l_2}^{{\cal A}}
(\gamma_{l_1,s_1}-\frac{\pi i}{n}|\{\alpha'\})
\Omega_{l_1,l_2}^{{\cal A}}
(\gamma_{l_1,s_1}+\frac{\pi i}{n}|\{\alpha'\})
}{
\Omega_{l_1,l_2}^{{\cal G}}(
\gamma_{l_1,s_1}|\{\gamma'\})
\Omega_{l_1,l_2}^{{\cal G}}
(\gamma_{l_1,s_1}-\frac{\pi i}{n}|\{\gamma'-\frac{\pi i}{n}\})
},\label{def:F1}
\end{eqnarray}
where
\begin{eqnarray}
\Omega_{l_1,l_2}^{{\cal A}}(\beta |\{\alpha\})=
\frac{
\displaystyle
\prod_{r_2 \in {\cal A}_{l_2-1}}
e_{l_1,l_2}^{(1)}\left(\beta+\alpha_{l_2-1,r_2}
-\frac{\pi i}{n}\right)
\prod_{r_2 \in {\cal A}_{l_2+1}}
e_{l_1,l_2}^{(1)}
\left(\beta+\alpha_{l_2+1,r_2}
-\frac{\pi i}{n}\right)}
{\displaystyle
\prod_{r_2 \in {\cal A}_{l_2}}
e_{l_1,l_2}^{(1)}\left(\beta
+\alpha_{l_2,r_2}\right)~
e_{l_1,l_2}^{(1)}
\left(\beta+\alpha_{l_2,r_2}
-\frac{2\pi i}{n}\right)}.\nonumber\\
\end{eqnarray}
Here we have set
\begin{eqnarray}
e_{l_1,l_2}^{(1)}(\alpha)=
\frac{\Gamma_3\left(
\left.i\alpha+\frac{\pi}{n}(l_1+l_2+2)
\right|
2\pi,2\pi,\frac{2\pi}{n}\right)
\Gamma_3\left(
\left.i\alpha-\frac{\pi}{n}(l_1+l_2-2)+2\pi
\right|
2\pi,2\pi,\frac{2\pi}{n}\right)
}{
\Gamma_3\left(
\left.i\alpha+\frac{\pi}{n}(l_2-l_1+2)
\right|
2\pi,2\pi,\frac{2\pi}{n}\right)
\Gamma_3\left(
\left.i\alpha+\frac{\pi}{n}(l_1-l_2+2)+2\pi
\right|
2\pi,2\pi,\frac{2\pi}{n}\right)
}.\nonumber\\
\end{eqnarray}

~\\
We have set
\begin{eqnarray}
\digamma_{l_1,l_2,l_3}^{(2)}(\{\alpha\},\{\gamma\})
&=&
\frac{\displaystyle
\prod_{r \in {\cal A}_{l_3-1}}
e_{l_1,l_2}^{(2)}\left(\alpha_{l_3-1,r}-\frac{\pi i}{n}\right)
\prod_{r \in {\cal A}_{l_3+1}}
e_{l_1,l_2}^{(2)}\left(\alpha_{l_3+1,r}-\frac{\pi i}{n}\right)
}{
\displaystyle
\prod_{r \in {\cal A}_{l_3}}
e_{l_1,l_2}^{(2)}(\alpha_{l_3,r})~
e_{l_1,l_2}^{(2)}\left(\alpha_{l_3,r}-\frac{2\pi i}{n}\right)
}\nonumber\\
&\times&
\frac
{\displaystyle
\prod_{s \in {\cal G}_{l_3}}
e_{l_1,l_2}^{(2)}\left(
\gamma_{l_3,s}+\frac{\pi i}{n}\right)
e_{l_1,l_2}^{(2)}\left(
\gamma_{l_3,s}-\frac{\pi i}{n}\right)
}
{\displaystyle
\prod_{s \in {\cal G}_{l_3-1}}
e_{l_1,l_2}^{(2)}\left(\gamma_{l_3-1,s}\right)
\prod_{s \in {\cal G}_{l_3+1}}
e_{l_1,l_2}^{(2)}\left(\gamma_{l_3+1,s}\right)
},\label{def:F2}
\end{eqnarray}
where we set
\begin{eqnarray}
e_{l_1,l_2}^{(2)}(\alpha)&=&
\sqrt{
\frac{\Gamma_3(
\left.i\alpha+\frac{\pi}{n}(l_1+l_2+1)
\right|\pi,2\pi,\frac{2\pi}{n})
\Gamma_3(
\left.i\alpha+2\pi+
\frac{\pi}{n}(-l_1-l_2+1)
\right|\pi,2\pi,\frac{2\pi}{n})
}{
\Gamma_3(
\left.i\alpha+\frac{\pi}{n}(l_1+l_2+2)
\right|\pi,2\pi,\frac{2\pi}{n})
\Gamma_3(
\left.i\alpha+2\pi+
\frac{\pi}{n}(-l_1-l_2+2)
\right|\pi,2\pi,\frac{2\pi}{n})
}}\nonumber\\
&\times&
\sqrt{
\frac{\Gamma_3(
\left.i\alpha+\frac{\pi}{n}(-l_1+l_2+2)
\right|\pi,2\pi,\frac{2\pi}{n})
\Gamma_3(
\left.i\alpha+2\pi+
\frac{\pi}{n}(l_1-l_2+2)
\right|\pi,2\pi,\frac{2\pi}{n})
}{
\Gamma_3(
\left.i\alpha+\frac{\pi}{n}(-l_1+l_2+1)
\right|\pi,2\pi,\frac{2\pi}{n})
\Gamma_3(
\left.i\alpha+2\pi+
\frac{\pi}{n}(l_1-l_2+1)
\right|\pi,2\pi,\frac{2\pi}{n})
}}.\nonumber
\\
\end{eqnarray}

~\\
We have set
\begin{eqnarray}
&&J_B(\{\alpha\}|\{\gamma\})_{j_1 \cdots j_N}^{k_1 \cdots k_M}
\nonumber\\
&=&
\prod_{l=1}^L
\frac{\digamma_{l,L,l}^{(3)}(
\{\alpha\}|\{\gamma\}|\mu+\frac{\pi i}{n}(2M-L))}{
\digamma_{l,L,l}^{(3)}(\{\alpha\}|\{\gamma\}|
\mu-\frac{\pi i}{n}L+2\pi i)}
\prod_{l=L+1}^{n-1}
\frac{\digamma_{L,l,l}^{(3)}(
\{\alpha\}|\{\gamma\}|\mu+\frac{\pi i}{n}(2M-L))}{
\digamma_{L,l,l}^{(3)}(\{\alpha\}|\{\gamma\}|
\mu-\frac{\pi i}{n}L+2\pi i)}\nonumber\\
&\times&
\prod_{l=1}^L
\frac{\digamma_{l,L,l}^{(3)}(
\{-\alpha\}|\{-\gamma\}|\mu+\frac{\pi i}{n}(2M-L))}{
\digamma_{l,L,l}^{(3)}(\{-\alpha\}|\{-\gamma\}|
\mu-\frac{\pi i}{n}L)}
\prod_{l=L+1}^{n-1}
\frac{\digamma_{L,l,l}^{(3)}(
\{-\alpha\}|\{-\gamma\}|\mu+\frac{\pi i}{n}(2M-L))}{
\digamma_{L,l,l}^{(3)}(\{-\alpha\}|\{-\gamma\}|
\mu-\frac{\pi i}{n}L)}
\nonumber\\
&\times&
\prod_{m=1}^M
\frac{\digamma_{m,M,m}^{(3)}(
\{\alpha\}|\{\gamma\}|-\mu-\frac{\pi i}{n}M+2\pi i)}{
\digamma_{m,M,m}^{(3)}(\{\alpha\}|\{\gamma\}|
-\mu+\frac{\pi i}{n}(2L-M)+2\pi i)}\nonumber\\
&\times&
\prod_{m=M+1}^{n-1}
\frac{\digamma_{M,m,m}^{(3)}(
\{\alpha\}|\{\gamma\}|-\mu-\frac{\pi i}{n}M+2\pi i)}{
\digamma_{M,m,m}^{(3)}(\{\alpha\}|\{\gamma\}|
-\mu+\frac{\pi i}{n}(2L-M)+2\pi i)}
\label{def:JB}\\
&\times&
\prod_{m=1}^M
\frac{\digamma_{m,M,m}^{(3)}(
\{-\alpha\}|\{-\gamma\}|-\mu-\frac{\pi i}{n}M+2\pi i)}{
\digamma_{m,M,m}^{(3)}(\{-\alpha\}|\{-\gamma\}|
-\mu+\frac{\pi i}{n}(2L-M))}
\prod_{m=M+1}^{n-1}
\frac{\digamma_{M,m,m}^{(3)}(
\{-\alpha\}|\{-\gamma\}|-\mu-\frac{\pi i}{n}M+2\pi i)}{
\digamma_{M,m,m}^{(3)}(\{-\alpha\}|\{-\gamma\}|
-\mu+\frac{\pi i}{n}(2L-M))},\nonumber
\end{eqnarray}
where we have set
\begin{eqnarray}
&&\digamma_{l_1,l_2,l_3}^{(3)}
(\{\alpha\}|\{\gamma\}|\nu)\nonumber\\
&=&
\frac{
\prod_{r \in {\cal A}_{l_3-1}}
e_{l_1,l_2}^{(3)}(
\alpha_{l_3-1,r}+\frac{\pi i}{n}-\nu)
\prod_{r \in {\cal A}_{l_3+1}}
e_{l_1,l_2}^{(3)}(
\alpha_{l_3+1,r}+\frac{\pi i}{n}-\nu)}{
\prod_{r \in {\cal A}_{l_3}}
e_{l_1,l_2}^{(3)}(\alpha_{l_3,r}-\nu)
e_{l_1,l_2}^{(3)}(\alpha_{l_3,r}+\frac{2\pi i}{n}-\nu)
}\nonumber\\
&\times&
\frac{
\prod_{s \in {\cal G}_{l_3}}
e_{l_1,l_2}^{(3)}(\gamma_{l_3,s}-\frac{\pi i}{n}-\nu)
e_{l_1,l_2}^{(3)}(\gamma_{l_3,s}+\frac{\pi i}{n}-\nu)
}{
\prod_{s \in {\cal G}_{l_3-1}}
e_{l_1,l_2}^{(3)}(\gamma_{l_3-1,s}-\nu)
\prod_{s \in {\cal G}_{l_3+1}}
e_{l_1,l_2}^{(3)}(\gamma_{l_3+1,s}-\nu)
}.\label{def:F3}
\end{eqnarray}
Here we have set
\begin{eqnarray}
e_{l_1,l_2}^{(3)}(\alpha)=
\frac{\Gamma_3(i\alpha+\frac{\pi}{n}(l_1+l_2+2)
|2\pi,2\pi,\frac{2\pi}{n})
\Gamma_3(i\alpha+2\pi+\frac{\pi}{n}(-l_1-l_2+2)
|2\pi,2\pi,\frac{2\pi}{n})
}{
\Gamma_3(i\alpha+\frac{\pi}{n}(-l_1+l_2+2)
|2\pi,2\pi,\frac{2\pi}{n})
\Gamma_3(i\alpha+2\pi+\frac{\pi}{n}(l_1-l_2+2)
|2\pi,2\pi,\frac{2\pi}{n})
}.\nonumber\\
\end{eqnarray}

~\\
{\it Let us summarize the result of this section.
We present integral representations 
(\ref{def:integral})
for the form factors
of the local operators (\ref{def:form}),
\begin{eqnarray}
f(\delta_1, \cdots, \delta_M|\beta_1,
\cdots, \beta_N)_{j_1 \cdots j_N}
^{k_1 \cdots k_M}.\nonumber
\end{eqnarray}
Here the factor $E(\{\beta\}|\{\delta\})$ is given in 
(\ref{def:E}).
The integrand $I(\{\alpha\}|\{\gamma\})_{j_1 \cdots j_N}
^{k_1 \cdots k_M}$ is given in (\ref{def:I}),
where the factor $J(\{\alpha\}|\{\gamma\})_{j_1 \cdots j_N}
^{k_1 \cdots k_M}
$,
$
J_F(\{\alpha\}|\{\gamma\})_{j_1 \cdots j_N}^{k_1 \cdots k_M}
$,
and
$
J_B(\{\alpha\}|\{\gamma\})_{j_1 \cdots j_N}
^{k_1 \cdots k_M}
$,
are
given in (\ref{def:J}),
(\ref{def:JF}), and (\ref{def:JB}).
Here the auxiliary functions
$
\digamma_{l_1,l_2}^{(1)}$,$
\digamma_{l_1,l_2,l_3}^{(2)}$,
$
\digamma_{l_1,l_2,l_3}^{(3)}$
are given by (\ref{def:F1}), (\ref{def:F2}),
and (\ref{def:F3}), respectively.
}

~\\
{\bf Acknowledgements}~~This work was partly supported by
Grant-in-Aid for Encouragements for Young Scientists ({\bf A})
from Japan Society for the Promotion of Science (11740099).

\begin{appendix}

\section{Multi-Gamma functions}

Here we summarize the multiple gamma and the multiple sine
functions.
Let us set the functions
$\Gamma_1(x|\omega), \Gamma_2(x|\omega_1, \omega_2)
$ and
$\Gamma_3(x|\omega_1, \omega_2, \omega_3)
$
by
\begin{eqnarray}
{\rm log}\Gamma_1(x|\omega)+\gamma B_{11}(x|\omega)&=&
\int_C\frac{dt}{2\pi i t}e^{-xt}
\frac{{\rm log}(-t)}{1-e^{-\omega t}},\\
{\rm log}\Gamma_2(x|\omega_1, \omega_2)
-\frac{\gamma}{2} B_{22}(x|\omega_1, \omega_2)&=&
\int_C\frac{dt}{2\pi i t}e^{-xt}
\frac{{\rm log}(-t)}{(1-e^{-\omega_1 t})
(1-e^{-\omega_2 t})},\\
{\rm log}\Gamma_3(x|\omega_1, \omega_2, \omega_3)
+\frac{\gamma}{3!} B_{33}(x|\omega_1, \omega_2, \omega_3)&=&
\int_C\frac{dt}{2\pi i t}e^{-xt}
\frac{{\rm log}(-t)}{(1-e^{-\omega_1 t})
(1-e^{-\omega_2 t})
(1-e^{-\omega_3 t})},\nonumber\\
\end{eqnarray}
where
the functions $B_{jj}(x)$ are the multiple Bernoulli polynomials
defined by
\begin{eqnarray}
\frac{t^r e^{xt}}{
\prod_{j=1}^r (e^{\omega_j t}-1)}=
\sum_{n=0}^\infty
\frac{t^n}{n!}B_{r,n}(x|\omega_1 \cdots \omega_r),
\end{eqnarray}
more explicitly
\begin{eqnarray}
B_{11}(x|\omega)&=&\frac{x}{\omega}-\frac{1}{2},\\
B_{22}(x|\omega)&=&\frac{x^2}{\omega_1 \omega_2}
-\left(\frac{1}{\omega_1}+\frac{1}{\omega_2}\right)x
+\frac{1}{2}+\frac{1}{6}\left(\frac{\omega_1}{\omega_2}
+\frac{\omega_2}{\omega_1}\right).
\end{eqnarray}
Here $\gamma$ is Euler's constant,
$\gamma=\lim_{n\to \infty}
(1+\frac{1}{2}+\frac{1}{3}+\cdots+\frac{1}{n}-{\rm log}n)$.\\
Here the contor of integral is given by

~\\
~\\

%WinTpicVersion2.13
\unitlength 0.1in
\begin{picture}(34.10,11.35)(17.90,-19.35)
% VECTOR 2 0 3 0
% 4 5200 1200 2190 1200 2190 2000 5190 2000
% 
\special{pn 8}%
\special{pa 5200 800}%
\special{pa 2190 800}%
\special{fp}%
\special{sh 1}%
\special{pa 2190 800}%
\special{pa 2257 820}%
\special{pa 2243 800}%
\special{pa 2257 780}%
\special{pa 2190 800}%
\special{fp}%
\special{pa 2190 1600}%
\special{pa 5190 1600}%
\special{fp}%
\special{sh 1}%
\special{pa 5190 1600}%
\special{pa 5123 1580}%
\special{pa 5137 1600}%
\special{pa 5123 1620}%
\special{pa 5190 1600}%
\special{fp}%
% LINE 2 0 3 0
% 2 5190 1600 2590 1610
% 
\special{pn 8}%
\special{pa 5190 1200}%
\special{pa 2590 1210}%
\special{fp}%
% STR 2 0 3 0
% 3 2590 1510 2590 1610 2 0
% $0$
\put(25.9000,-12.1000){\makebox(0,0)[lb]{$0$}}%
% CIRCLE 2 0 3 0
% 4 2190 1610 2190 1210 2190 1210 2190 4410
% 
\special{pn 8}%
\special{ar 2190 1210 400 400  1.5707963 4.7123890}%
% STR 2 0 3 0
% 3 3390 2320 3390 2420 5 0
% {\bf Contour} $C$
\put(33.9000,-20.2000){\makebox(0,0){{\bf Contour} $C$}}%
\end{picture}%

~\\
~\\

Let us set
\begin{eqnarray}
S_1(x|\omega)&=&\frac{1}{\Gamma_1(\omega-x|\omega)
\Gamma_1(x|\omega)},\\
S_2(x|\omega_1,\omega_2)&=&\frac{
\Gamma_2(\omega_1+\omega_2-x|\omega_1,\omega_2)}{
\Gamma_2(x|\omega_1,\omega_2)},\\
S_3(x|\omega_1,\omega_2,\omega_3)&=&\frac{1}{
\Gamma_3(\omega_1+\omega_2+\omega_3-x|\omega_1,\omega_2,\omega_3)
\Gamma_3(x|\omega_1,\omega_2,\omega_3)}
\end{eqnarray}
We have
\begin{eqnarray}
\Gamma_1(x|\omega)=e^{(\frac{x}{\omega}-\frac{1}{2}){\rm log}
\omega}\frac{\Gamma(x/\omega)}{\sqrt{2\pi}},~
S_1(x|\omega)=2{\rm sin}(\pi x/\omega),
\end{eqnarray}
\begin{eqnarray}
\frac{\Gamma_2(x+\omega_1|\omega_1,\omega_2)}{
\Gamma_2(x|\omega_1,\omega_2)}=\frac{1}{\Gamma_1(x|\omega_2)},~
\frac{S_2(x+\omega_1|\omega_1,\omega_2)}{
S_2(x|\omega_1,\omega_2)}=\frac{1}{S_1(x|\omega_2)},~
\frac{\Gamma_1(x+\omega|\omega)}{\Gamma_1(x|\omega)}=x.
\end{eqnarray}

\begin{eqnarray}
\frac{
\Gamma_3(x+\omega_1|\omega_1,\omega_2, \omega_3
}{\Gamma_3(x|\omega_1,\omega_2, \omega_3)}
=\frac{1}{\Gamma_2(x|\omega_2, \omega_3)},~
\frac{S_3(x+\omega_1|\omega_1,\omega_2,\omega_3)}{
S_3(x|\omega_1,\omega_2,\omega_3)}=\frac{1}{S_2(x|\omega_2,
\omega_3)}.
\end{eqnarray}

\begin{eqnarray}
{\rm log}S_2(x|\omega_1 \omega_2)
=\int_C \frac{{\rm sh}(x-\frac{\omega_1+\omega_2}{2})t}
{2{\rm sh}\frac{\omega_1 t}{2}
{\rm sh}\frac{\omega_2 t}{2}
}{\rm log}(-t)\frac{dt}{2\pi i t},~(0<{\rm Re}x<
\omega_1+\omega_2).
\end{eqnarray}

\begin{eqnarray}
S_2(x|\omega_1 \omega_2)=
\frac{2\pi}{\sqrt{\omega_1 \omega_2}}x +O(x^2),~~(x \to 0).
\end{eqnarray}

\begin{eqnarray}
S_2(x|\omega_1 \omega_2)
S_2(-x|\omega_1 \omega_2)=-4
{\rm sin}\frac{\pi x}{\omega_1}
{\rm sin}\frac{\pi x}{\omega_2}.
\end{eqnarray}

\end{appendix}
\end{document}